\documentclass[twocolumn,aps,pra,amsmath,amssymb,superscriptaddress,floatfix]{revtex4-2}

\usepackage{color}
\usepackage{graphicx}
\usepackage{xspace} 
\usepackage{dcolumn}
\usepackage{bm}

\begin{document}

\title{Quasiparticle spectroscopy, transport, and magnetic properties\\ of Nb films used in superconducting transmon qubits}

\author{Kamal R. Joshi}
\affiliation{Ames National Laboratory, Ames, IA 50011, U.S.A.}

\author{Sunil Ghimire}
\affiliation{Ames National Laboratory, Ames, IA 50011, U.S.A.}
\affiliation{Department of Physics \& Astronomy, Iowa State University, Ames, IA 50011, U.S.A.}

\author{Makariy A. Tanatar}
\affiliation{Ames National Laboratory, Ames, IA 50011, U.S.A.}
\affiliation{Department of Physics \& Astronomy, Iowa State University, Ames, IA 50011, U.S.A.}

\author{Amlan Datta}
\affiliation{Ames National Laboratory, Ames, IA 50011, U.S.A.}
\affiliation{Department of Physics \& Astronomy, Iowa State University, Ames, IA 50011, U.S.A.}

\author{Jin-Su Oh}
\affiliation{Ames National Laboratory, Ames, IA 50011, U.S.A.}

\author{Lin Zhou}
\affiliation{Ames National Laboratory, Ames, IA 50011, U.S.A.}

\author{Cameron J. Kopas}
\affiliation{Rigetti Computing, 775 Heinz Ave., Berkeley, CA 94710, U.S.A.}

\author{Jayss Marshall}
\affiliation{Rigetti Computing, 775 Heinz Ave., Berkeley, CA 94710, U.S.A.}

\author{Josh Y. Mutus}
\affiliation{Rigetti Computing, 775 Heinz Ave., Berkeley, CA 94710, U.S.A.}

\author{Julie Slaughter}
\affiliation{Ames National Laboratory, Ames, IA 50011, U.S.A.}

\author{Matthew J. Kramer}
\affiliation{Ames National Laboratory, Ames, IA 50011, U.S.A.}

\author{James A. Sauls}
\affiliation{Center for Applied Physics and Superconducting Technologies, Department
of Physics and Astronomy Northwestern University, Evanston, IL 60208, U.S.A.}

\author{Ruslan Prozorov}
\email[Corresponding author: ]{prozorov@ameslab.gov}
\affiliation{Ames National Laboratory, Ames, IA 50011, U.S.A.}
\affiliation{Department of Physics \& Astronomy, Iowa State University, Ames, IA 50011, U.S.A.}

\date{23 July 2022}

\begin{abstract}
Niobium thin films on silicon substrate used in the fabrication of
superconducting qubits have been characterized using scanning and transmission
electron microscopy, electrical transport, magnetization, quasiparticle
spectroscopy, and real-space real-time magneto-optical imaging. We study niobium films to provide an example of a comprehensive analytical set that may benefit superconducting circuits such as those used in quantum computers. The
films show outstanding superconducting transition temperature of $T_{c}=9.35$
K and a fairly clean superconducting gap, along with superfluid density enhanced at intermediate temperatures. These observations are consistent with the recent theory of anisotropic strong-coupling superconductivity in Nb. However, the
response to the magnetic field is complicated, exhibiting significantly
irreversible behavior and insufficient heat conductance leading to
thermo-magnetic instabilities. These may present an issue for further
improvement of transmon quantum coherence. Possible mitigation strategies
are discussed.
\end{abstract}

\maketitle

\section{Introduction}

Superconducting qubits are promising candidates for implementing large-scale
quantum computers \cite{Reagor2016,Kjaergaard2020,Alam2022}. The
advancement in fabrication and design of superconducting qubits have
demonstrated impressive gate fidelity of up to 99.5 \% for two-qubit
gates, which is the measure of the ability of a device to faithfully execute quantum algorithms
\cite{Jurcevic2021}. However, large-scale devices will require fidelity
well beyond 99.9 \% \cite{Leon2021}. It was shown that certain impurities
and defects in qubit material tend to shorten the coherence time and lead to overall lower fidelity operations \cite{Schloer2019,Lee2021,Murthy2022}.
For example, microwave loss at cryogenic temperatures has been attributed
to the resonant coupling to two-level systems (TLS) physically composed
of the atomic defects \cite{Mueller2019,Lee2021}. Therefore, achieving
better fidelities requires improved quantum coherence and that necessiates a  better understanding and control over the materials used to fabricate qubits.

Niobium is the material of choice for various superconducting applications
due to its relatively high transition temperature, $T_{c}=9.35$
K. Niobium thin films (100-200 nm thick) are used in the superconducting
transmon qubits \cite{Reagor2016}. The ``heart'' of such qubits
is an aluminum Josephson junction used to provide non-linearity to the circuit and niobium is used to fabricate the vast majority of the rest of the circuit:
the readout resonators, capacitor pads, and coupling lines in the transmon.
The structure must have a high degree of quantum coherence in all parts
to work. Surfaces and interfaces are known to introduce noise and
loss that play a major role in decreasing the coherence time in transmon
qubits \cite{Oliver2013,Wang2015,Dial2016,Gambetta2017,Lee2021}.
For instance, the presence of native amorphous oxides with different
stoichiometry (NbO, NbO$_{2}$ and Nb$_{2}$O$_{5}$) are believed
to be the host for TLSs, hence decoherence \cite{Burnett2016,Niepce2020,Verjauw2021,Premkumar2021}.
Another recently discovered source of potential decoherence in transmons
is nano-sized niobium hydrides \cite{Lee2021} (large hydrides in
superconducting RF cavities are believed to be the cause of the so-called
Q-disease (a dramatic reduction of the quality factor above some amplitude of electromagnetic field inside)
and have been known for a long time \cite{Anderson1959a,Arai2004,Bahte1998,Barkov2013,Barkov2012,Dhakal2017}).
While bulk niobium used in SRF cavities was extensively characterized
both in normal and the superconducting states, the niobium films used
in QIS technologies have mostly been studied in the normal state and, in the superconducting state, specifically to examine the quality factor behavior and losses at GHz frequencies \cite{ValenteFeliciano2022}. Bulk and film states of niobium are very different in terms of morphology, purity, and relative length scales involved, and more conventional studies in the superconducting state are needed.

In this work, we employ various techniques to study thermally excited
quasiparticles, electrical transport, upper and third critical fields,
magnetization and spatial distribution of the magnetic induction in
niobium thin films and suggest the strategies for further optimizations
and improvements.

\section{EXPERIMENTAL}

Niobium films, 160 nm thick, were deposited onto {[}001{]} high-resistivity silicon wafers ($> 10000$ $\Omega\cdot cm$
using high-power impulse magnetron sputtering (HiPIMS) in a ultra-high vacuum system with a base pressure $<1\times 10^{-8}$ Torr. \cite{Murthy2022}.

The film morphology was studied using scanning and transmission electron
microscopy, high-resolution SEM, and TEM. A more detailed investigation
of the transmons from the same batch as in this work, and imaging
details, are available elsewhere \cite{Oh2022}. Cross-sectional and plan-view TEM samples were prepared with a Helios focused ion beam system. The TEM images were acquired at an acceleration voltage of 200 kV using a Titan Themis. The SEM was used to map out the larger view of sample surface morphology. The TEM images were used for quantitative analysis.

To study the low-energy quasiparticles, we used a tunnel-diode resonator
for the precision measurements of the London penetration depth, $\Delta\lambda(T)$
\cite{Prozorov2006,Prozorov2011}, which is then used to calculate
the superfluid density, which is compared to the expectations from
the theory.

Four-probe electrical resistivity measurements were performed in a \emph{Quantum
Design} PPMS. Contacts were made by gluing 25 $\ensuremath{\mathrm{\mu m}}$
silver wires using DuPont 4929N conducting silver paste. This technique
yields contacts with contact resistance in the 10 to 100 $\Omega$ range.
Transport measurements of the upper critical field, $\ensuremath{H_{c2}}$,
were performed with a magnetic field oriented perpendicular to the
film plane to avoid the third critical field, $\ensuremath{H_{c3}}$,
which does not exist in this orientation but is maximum when the magnetic
field is parallel to the film surface \cite{Abrikosov2017,Kogan2002}.
For measurements in a parallel configuration, the sample was glued on the side of the plastic cube, similar to those described in Ref.~\onlinecite{YongLiu}. This procedure provided alignment with the accuracy of about 2 degrees.

Total magnetization was measured using a vibrating sample magnetometer
(VSM) module in \emph{Quantum Design} PPMS. The advantage of this
device is the ability to sweep the magnetic field continuously at a constant
rate.

The two-dimensional distribution of the magnetic induction was mapped
in real-time employing magneto-optical imaging using Faraday Effect
(MOFA) in transparent ferrimagnetic indicators (bismuth-doped iron
garnets) placed on top of the samples. The closed-cycle flow-type
optical $^{4}$He cryostat exposed the cooled sample to an Olympus
polarized-light microscope. The magnetic induction on the sample surface
polarizes in-plane magnetic moments in the indicator, and the distribution
of this polarization component along the light propagation is visualized
through double Faraday rotation. In the images, only the magnetic field
is visible due to a mirror sputtered at the bottom of the indicator
\cite{Indenbom1993,Prozorov2005,Prozorov2007,Prozorov2008}.

\section{Results}

\begin{figure}[tb]
\centering \includegraphics[width=8.5cm]{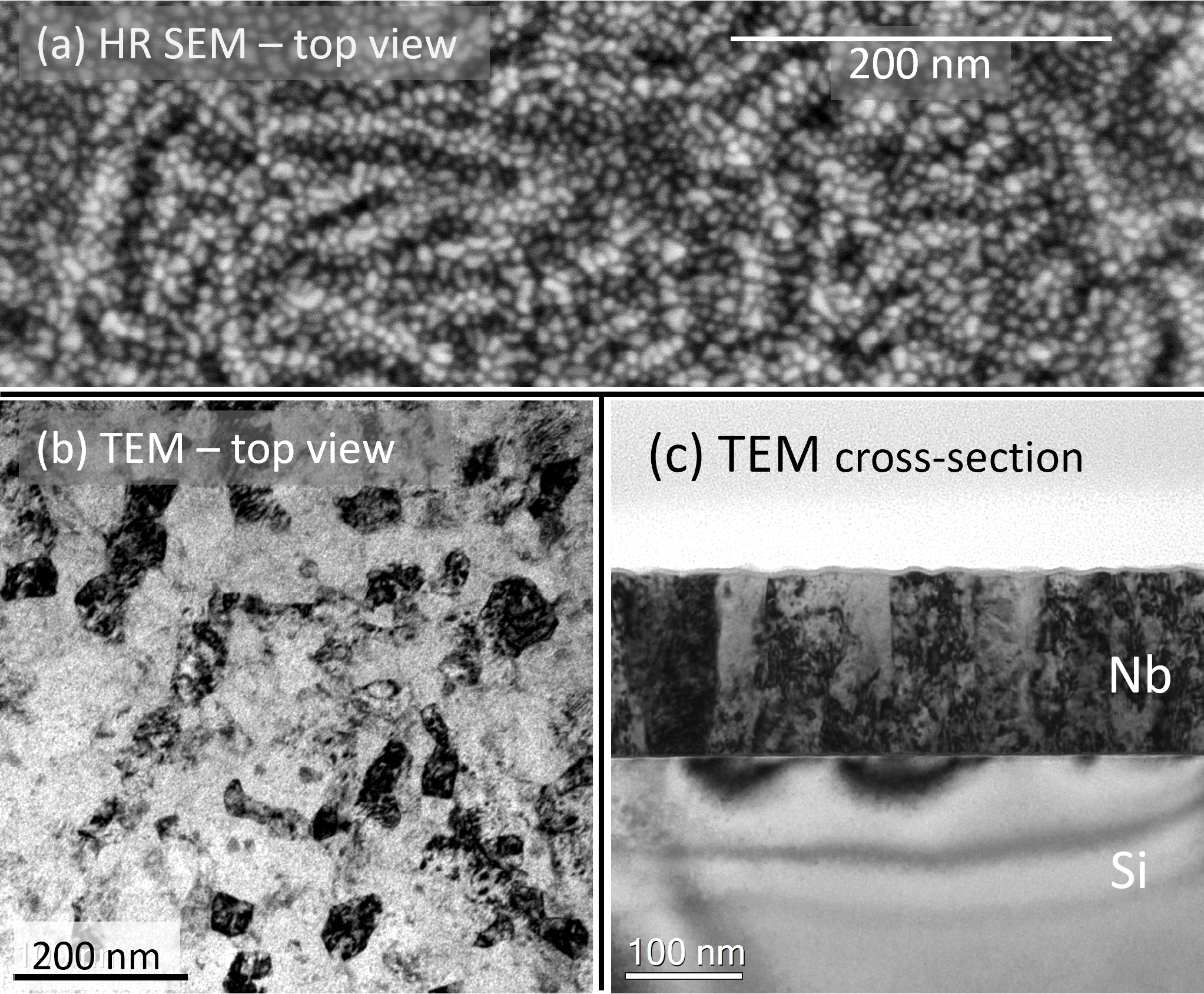}
\caption{(a) Secondary electrons high-resolution SEM wide area view of sample surface morphology. (b) TEM plan view and (c) TEM cross-sectional view of the studied Nb thin film. The bulk consists of columnar Nb grains with sizes ranging between 25 and 75 nm, with an average grain size of about 50 nm. The surface roughness is of the order of 3 to 5 nm. }
\label{fig1}
\end{figure}

Figure \ref{fig1} shows the morphology of the Nb film revealed by high-resolution scanning (HR-SEM) and transmission (TEM) electron microscopy. Figure \ref{fig1}(a) shows the SEM image of a wide view of the top surface morphology, revealing substantial spatial modulations at two characteristic length scales. Further details of internal and surface structural texture are revealed by bright-field plan-view TEM image, Fig.\ref{fig1}(b), and cross-sectional view, Fig.\ref{fig1}(c). The contrast variation reveals granular polycrystalline structure on top and inside the film with grain sizes ranging from 25 nm to 75 nm, with an average size of about 50 nm. Those grains have a columnar structure along film thickness, measured at 160 nm as intended by the fabrication. The smaller droplets observed in secondary electrons contrast, Fig.\ref{fig1}(a), represent surface roughness also seen in the cross-sectional view, Fig.\ref{fig1}(c).

\begin{figure}[tb]
\centering \includegraphics[width=8.5cm]{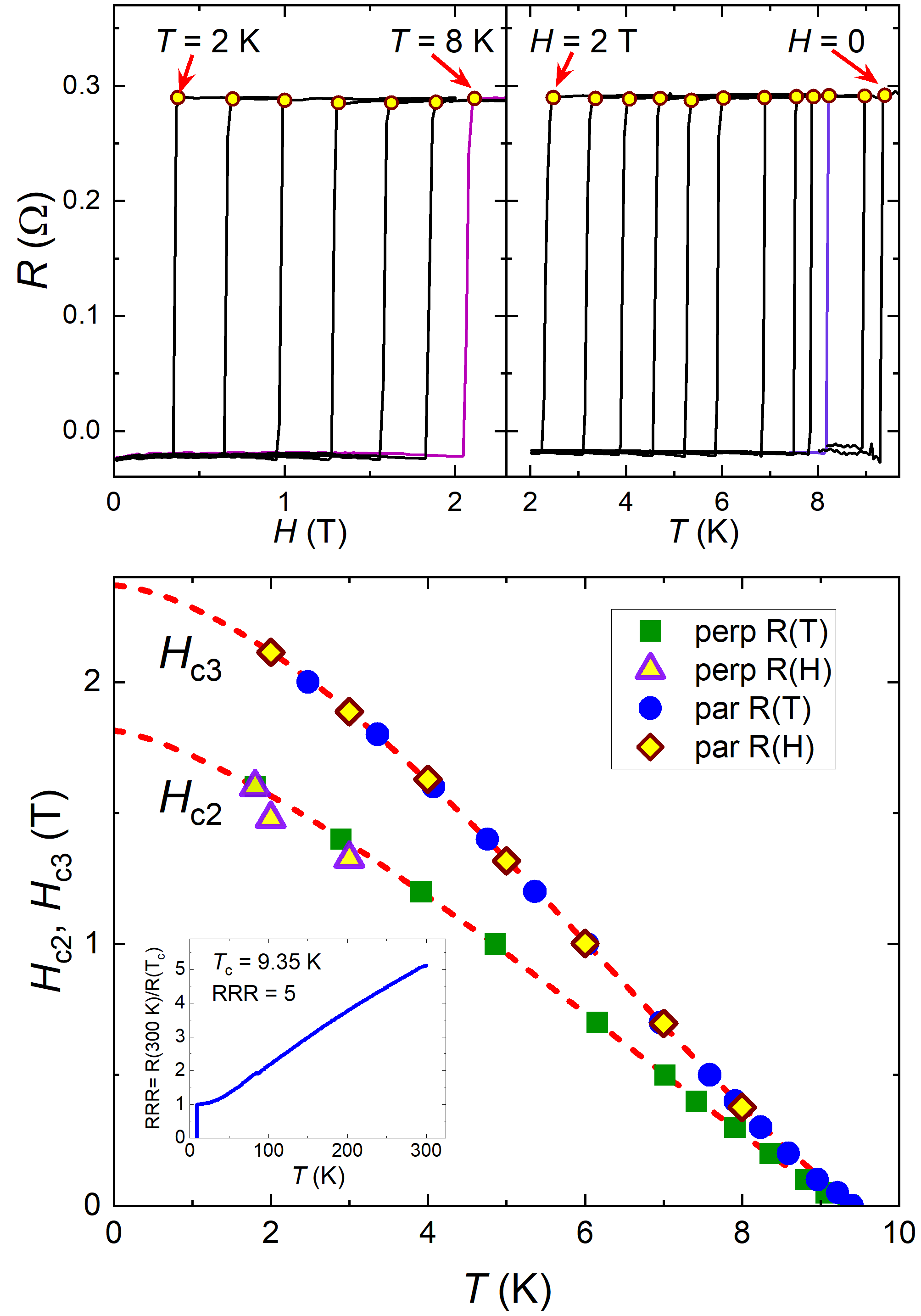}
\caption{Top panels: electrical resistivity of Nb film in magnetic field parallel to the film plane. (Left) $R\left(H\right)$
scans measured at different temperatures from 2 to 8 K with 1 K interval.
(Right) $R\left(T\right)$ scans at fixed magnetic fields of, from
right to left: $H=$0, 0.1, 0.3, 0.4, 0.5, 0.7, 1.0, 1.2, 1.4, 1.6,
1.8 and 2 T. Bottom panel: summary phase diagram showing the third
critical field, $H_{c3}$, obtained from the data shown in the top
panels, and the upper critical field, $H_{c2}$, obtained from the
similar measurements with magnetic field oriented perpendicular to
the plane. The inset shows temperature-dependent resistivity normalized by its value at $T_c$. The   $RRR=R(300\:\mathrm{K})/R\left(T_{c}\right)=5$.}
\label{fig2}
\end{figure}

Figure \ref{fig2} presents electrical resistivity measurements as a function
of magnetic field and temperature. The data were collected in parallel
and perpendicular orientations of the film with respect to the applied
magnetic field. In a parallel orientation, the material remains superconducting
up to a so-called third critical magnetic field, $H_{c3}$, surviving
in a thin surface later of the order of the superconducting coherence
length, $\xi$ \cite{Abrikosov2017,Kogan2002,Liarte_2017}. In niobium,
$\xi\approx10-40$ nm, depending on its purity and degree of crystallinity
\cite{Abrikosov2017,Finnemore1966,Lechner2020,Zarea2022}. In the
perpendicular orientation of the magnetic field to the film plane,
the upper critical field, $H_{c2}$, is the highest magnetic field
possible \cite{Abrikosov2017,Kogan2002,Xie2017}. Therefore, we obtain
these two critical magnetic fields from the measurements of the same
sample, simply rotating it by 90 degrees without changing the contacts.
In the PPMS, the cooling is achieved via the helium exchange gas when
the sample is mounted on a puck, and so the orientation is not important
for thermal equilibrium as long as the measurements are performed
slowly. Indeed, we see the same $T_{c}=9.35$ K in zero magnetic
field in both orientations.

The top left panel of Fig.\ref{fig2} shows the magnetic field - dependent
resistance, $R\left(H\right)$, measured at different temperatures
from 2 to 8 K in 1 K steps. The top right panel shows the temperature-dependent
resistance, $R\left(T\right)$, measured at different magnetic fields,
from right to left: $H=$0, 0.1, 0.3, 0.4, 0.5, 0.7, 1.0, 1.2, 1.4,
1.6, 1.8, and 2 T. In both top panels, the symbols show the locations
of the third critical field, $H_{c3}\left(T\right)$ (left panel),
or critical temperatures, $T_{c}\left(H_{c3}\right)$ (right panel),
an easy pick, considering how sharp the transitions remain for all
curves. Similar measurements were performed in the perpendicular orientation,
yielding the upper critical field, $H_{c2}\left(T\right)$. The bottom
panel of Fig.\ref{fig2} summarizes the results of the critical fields
measurements. The upper curve, obtained from the data in the top panels,
shows the third critical field, $H_{c3}\left(T\right)$. The agreement
between the two types of scans is outstanding. Similarly, the lower
curve shows the upper critical field, $H_{c2}\left(T\right)$.

The inset shows temperature-dependent resistivity normalized by its value at $T_c$ in a broad temperature range up to 300 K. At 300 K, this ratio, called the residual resistivity ratio, $RRR(300\:\mathrm{K})\approx5$, is the commonly used measure of scattering on defects and impurities. For comparison, the cleanest niobium samples reach $RRR=90000$ \cite{Koethe2000}, and so the value of 5 for our films is rather low.

\begin{figure}[tb]
\centering \includegraphics[width=8.5cm]{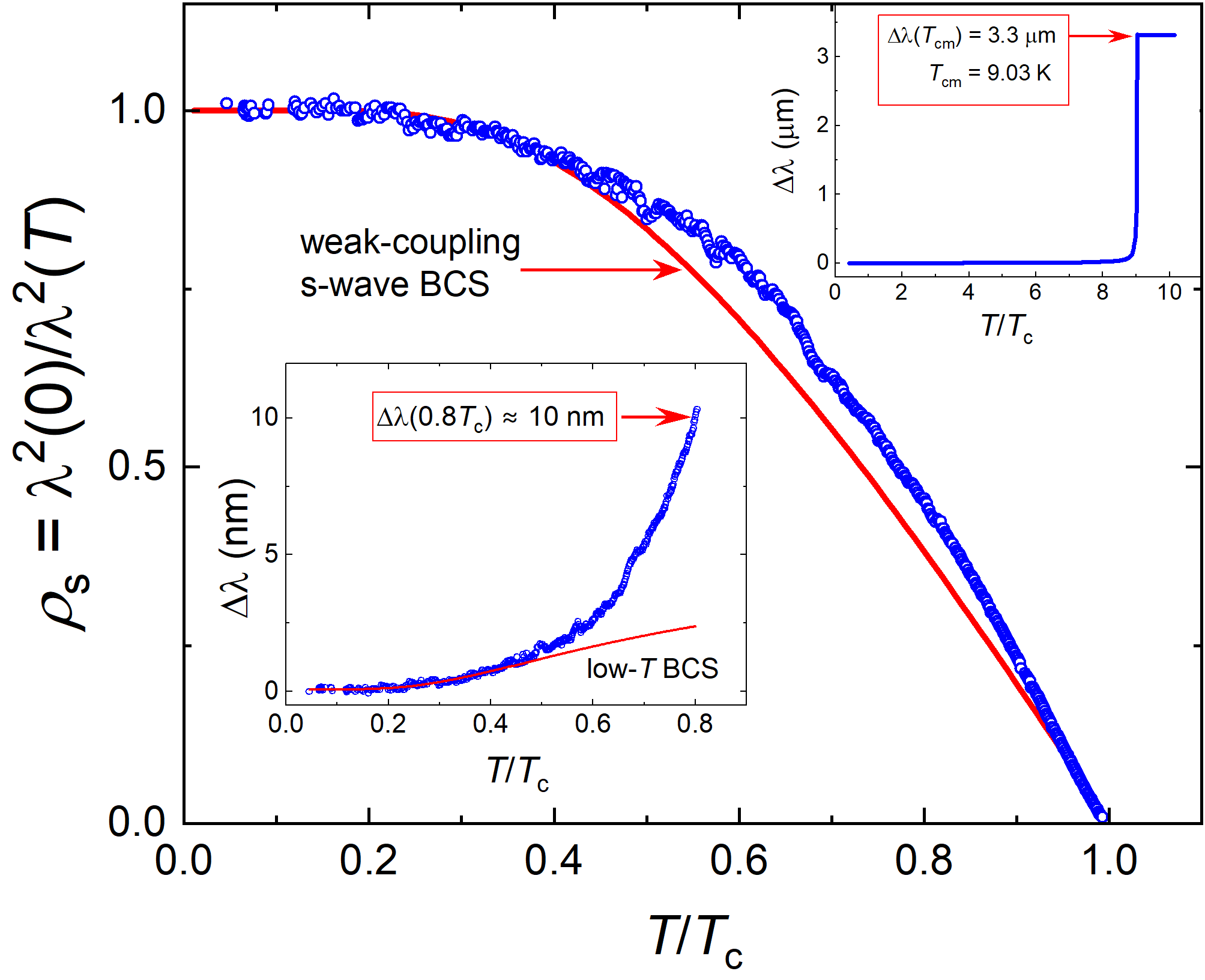}
\caption{(color online) Main panel: normalized superfluid density evaluated
from the measured London penetration depth shown in the insets using
the formula shown in the Y-axis title. The solid red curve is the
expectation from the isotropic weak-coupling BCS theory \cite{Bardeen1957}.
The upper inset shows a full-temperature range variation, whereas the lower
left inset shows the data up to $0.8T_{c}$. The red line is the fit
to the isotropic low-temperature asymptotic BCS formula described
in the text. Note very small total variation of $\Delta\lambda\left(T\right)$
due to significantly reduced effective sample dimension due to thin-film geometry \cite{Prozorov2021}.}
\label{fig3}
\end{figure}

We now examine the superconducting state of the studied films. Figure
\ref{fig3} shows the temperature-dependent superfluid density, $\rho\left(T\right)=\lambda^{2}\left(0\right)/\lambda^{2}\left(T\right)$,
obtained from the London penetration depth, $\lambda\left(T\right)$,
measured using tunnel-diode resonator \cite{Prozorov2006,Prozorov2011}.
The solid red curve is the expectation from the isotropic weak-coupling
Bardeen-Cooper-Schrieffer (BCS) theory \cite{Bardeen1957}. This measurement
is quite difficult due to the extreme thin-film geometry. The upper
inset in Fig.\ref{fig3} shows the full transition curve, which looks
flawless, but cuts off at about $\Delta\lambda\left(T_{cm}\right)=3.3\:\mathrm{\mu m}$,
at which point the film becomes transparent to our 10 MHz RF AC field.
This formally follows from the analysis of the magnetic susceptibility
in finite samples \cite{Prozorov2021}. For our films, the formal
calibration gives for the effective sample size, $R=3\:\mathrm{\mu m}$,
which matches perfectly the above experimental value for the penetration
depth at $T_{c}$ when the response becomes ``sample size limited''
\cite{Prozorov2021}. This also explains why the measured $T_{cm}=9.03$
K is a little lower than $T_{c}=9.35$ K detected by the transport
measurements, Fig.\ref{fig2}. The lower inset zooms into the lower
temperature behavior. Remarkably, the penetration depth changes only
by 10 nm from the base temperature up to $0.8T_{c}$. And yet we were
able to resolve the whole superfluid density curve shown in the main
panel of Fig.\ref{fig3}. A clear exponential attenuation
at low temperatures signals a full superconducting gap. If we
fit the data to a low-temperature BCS formula, valid below $0.3T_{c}$,
$\Delta\lambda/\lambda\left(0\right)=\sqrt{\pi\delta/2t}\exp\left(-\delta/t\right)$
\cite{Prozorov2011}, where $\delta=\Delta\left(0\right)/k_{B}T_{c}$
and $t=T/T_{c}$ and use $\delta$ as a free parameter, we obtain
$\delta\approx1.3$ instead of isotropic BCS value of $\delta\approx1.76$.
This is most likely associated with the electronic and superconducting
states anisotropy of niobium, and its multiband nature \cite{Zarea2022}.
In this situation, the range of exponential attenuation is determined
by the smaller gap, specifically on band 2, which also dominates the
thermodynamic response \cite{Zarea2022}. The theory is further confirmed
by the observed deviation of the experimental superfluid density at
elevated temperatures from the weak coupling expectation, shown by
the solid red line in the main panel of Fig.\ref{fig3}. Stronger
coupling makes $\rho\left(T\right)$ lie above the weak coupling values
\cite{Zarea2022}.

\begin{figure}[tb]
\centering \includegraphics[width=8.5cm]{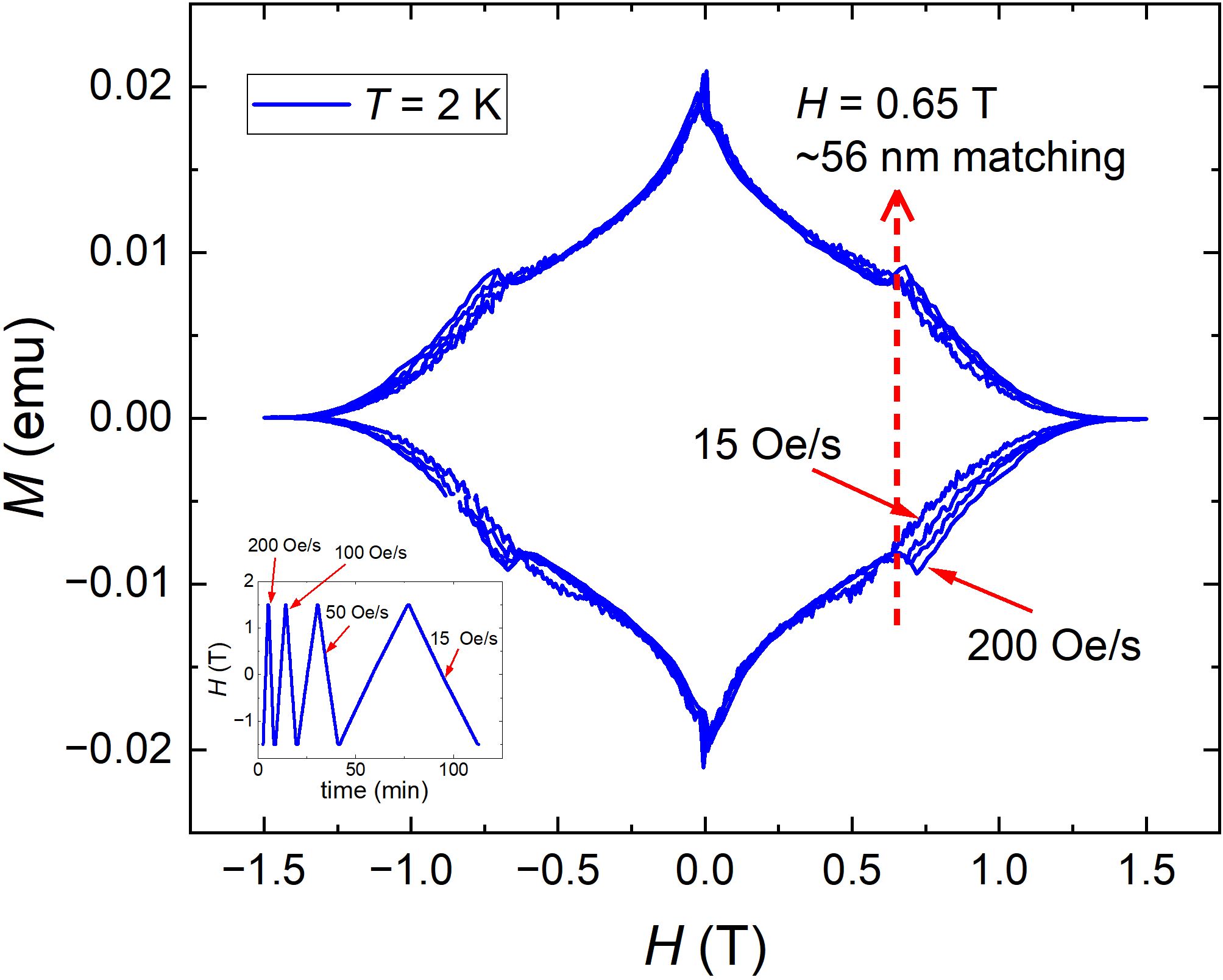}
\caption{The VSM measurements of the hysteresis loops at $T=2$ K while continuously
sweeping an applied magnetic field up and down four times, each time
at a different rate, to examine the dynamic effects. The field time
variation and the rate values are shown in the inset. The measurements
were performed in a magnetic field perpendicular to the film plane. Curiously, an additional
hysteretic response appears above the so-called ``matching field'' of
0.65 T corresponding roughly to 56 nm length scale matching the structural
features found by EM imaging, Fig.\ref{fig1}, see text for details.}
\label{fig4}
\end{figure}

Figure \ref{fig4} shows profound features on the $M(H)$ magnetic
hysteresis loops obtained using \emph{Quantum Design} PPMS-VSM. This
instrument allows for the measurements while sweeping the magnetic field
continuously without pausing. The field time variation is shown in
the inset with 200, 100, 50, and 15 Oe/s rates, more than
ten times the span of $dH/dt$. The measurements shown in \ref{fig4}
were performed perpendicular to the film plane. Figure \ref{fig4}
shows additional dynamic rate-dependent hysteretic response appearing
above roughly 0.65 T. In the simplest approximation, if the average
distance between Abrikosov vortices is $a$, it corresponds to the
magnetic induction of $B=\phi_{0}/a^{2}$, where $\phi=2.07\times10^{-15}$
Wb is the magnetic flux quantum. If there are some pinning centers
of a characteristic length-scale, $\ell_{p}$, the vortex pinning is
most efficient when the inter-vortex distance matches this scale, $a\simeq\ell_{p}$.
This defines the ``matching field'', $H_{m}$. Therefore, we can
estimate $\ell_{p}=\sqrt{\phi_{0}/H_{m}}\approx56$ nm, where,
in our case, $H_{m}=0.65$ T, Fig.\ref{fig4}. This immediately reminds of Fig.\ref{fig1}
where this length scale was determined from the direct EM imaging, providing
further evidence of the importance of the film morphology in determining
the magnetic response. Overall, our films show a much larger magnetic
hysteresis compared to the bulk samples and single crystals \cite{Finnemore1966,Stromberg1965}.
However, a very large demagnetizing factor and thin-film geometry
play a significant role in diminishing the interval and the magnitude
of the reversible magnetic moment, which is proportional to a very
small film volume, whereas irreversible Bean currents flow in the
entire sample \cite{Bean1962,Bean1964,Brandt1995,Brandt1998,Brandt1996}.

\begin{figure}[tb]
\centering \includegraphics[width=9cm]{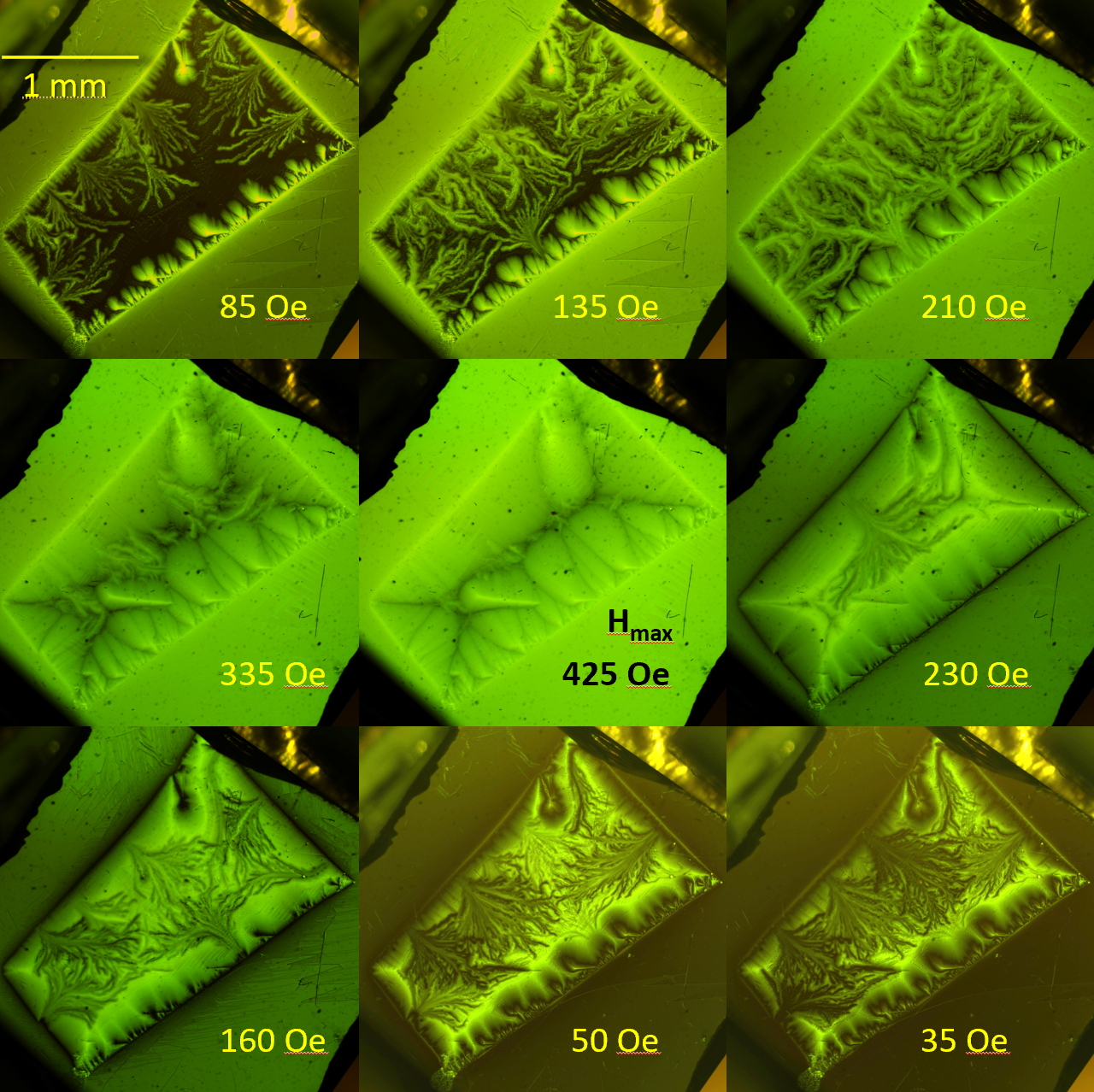}
\caption{Magneto-optical imaging of the magnetic induction in a thin film sample
at $T=5$ K. Magnetic field was applied perpendicular to the film plane after cooling the sample in
zero field. The first five images show flux penetration, $H=85, 135, 210, 335,$ and $425$ Oe. After that, the magnetic field was reduced imaging flux exit and trapping in the sample, $H=230, 160, 50,$ and $35$ Oe. The video of this process is available as supplementary material. The scale bar is shown in the top left image. }
\label{fig5}
\end{figure}

We now examine the details of the magnetic flux penetration and exit
using direct magneto-optical imaging of the superconducting state
using specially fabricated transparent ferrimagnetic indicators placed
on top of samples \cite{Indenbom1993,Young2005,Prozorov2006a}. Figure
\ref{fig5} shows the sequence of images obtained after cooling in
zero magnetic field to 5 K. The first five frames (left to right, top
to bottom) show increasing magnetic field up to the maximum of $H_{max}=425$
Oe, after which the magnetic field was reduced, and flux was exiting
the sample. There are two distinct regimes of flux penetration and
exit. First, large-scale, very fast (see the video in supplemental
information), dendritic avalanches similar in their appearance to lightning
strikes. Similar behavior was observed long ago in various films,
particularly niobium, and is well-understood \cite{BlancoAlvarez2019,Vestgaarden2018}.
The physics behind it is straightforward. When first vortices are pushed
inside by the screening currents, these moving vortices induce an
electric field parallel to the currents, resulting in a resistive
state, and energy is dissipated. If there is an insufficient thermal
link to the surrounding and, in the case of films, to the substrate, the
released heat decreases the local critical current in the direction of
the vortex motion creating a self-heating violent unidirectional propagation
of bunches of vortices that follow the heated up trail which needs
time to dissipate. Sometimes they branch out when meeting various
obstacles and, in this case, most likely the grain boundaries. In
thicker samples (a few tens of $\mu$m foils), the avalanches may become
``global'' leading to a catastrophic collapse of the entire Bean
critical state \cite{Prozorov2006a}. The actual local temperature
instabilities have been measured during these processes. In the case
of QIS applications, this insufficient heat channeling may represent
a difficult problem. The second stage of flux penetration is a regular
Bean critical state pattern \cite{Bean1962,Bean1964,Brandt1998},
also somewhat non-uniform, following the granular film morphology.
When the external magnetic field is decreased, the reverse process
happens, and similar dendritic avalanches start to channel out, now
shown by a darker color corresponding to the local suppression of the
magnetic induction. In the end, the mix of uniform and non-uniform
trapped flux remains in the sample as a fingerprint of the intense
perturbations occurring when the field was ramped up and down. The
supplemental information shows a real-time movie of the processes
pictured in Fig.\ref{fig5}. This highly inhomogeneous magnetic flux evolution reveals the issues of thermal heat-sinking and morphological
inhomogeneity issues.

\section{Discussion}

Let us review the experimental findings and try to understand all different
results from a common point of view. All measurements indicate that
as far as superconducting properties are concerned, the films are
rather far from being ideal and exhibit various irreversible properties
that generally lead to dissipation, hence quantum decoherence.

One of the most puzzling facts is that while the $RRR=5$ is rather
very low, the superconducting transition temperature, $T_{c}=9.35$
K, is as high as in good single crystals. According to the theory,
$T_{c}$ should be suppressed by non-magnetic disorder due to anisotropy
of the order parameter and a multi-band nature of the material \cite{Zarea2022}.
However, the same theory predicts saturation of the $T_{c}$ when
the gap averages out at higher scattering rates. Still, $T_{c}$ should
be lower than 9 K, and it isn't. One natural way to understand it,
supported by the direct EM imaging, Fig.\ref{fig1}, is to assume
that grains themselves are quite clean in terms of the scattering
rate, but the random network of grain boundaries creates significant
resistance. The boundaries do not affect high $T_{c}$ inside the
grains, though. In the normal state, just above $T_{c}$, the resistivity
is mostly determined by those boundaries. However, when the grains
become superconducting, the boundaries become a network of Josephson
junctions, and the whole sample transitions to zero resistance. This
explains the low $RRR$ and, simultaneously, high $T_{c}$.

The heterogeneous granular structure also explains the measurements
of the third and the upper critical fields, shown in Fig.\ref{fig2}.
While it is important to measure these two fields in one experiment
and on the same sample, the problem is that their ratio is about 1.3,
whereas a ratio of at least 1.7 is expected theoretically even in
the dirty limit \cite{Xie2017}. However, real sample geometry and
surface conditions may significantly affect the $H_{c3}$ \cite{Kogan2002}
beyond the point-like scattering and smooth semi-infinite surface
with perfectly parallel field considered in Ref.\cite{Xie2017}. Therefore,
our granular film does not show the expected maximum value of this
ratio. This is further re-enforced by the clear matching effect at
the length-scale of around 50 nm, Fig.\ref{fig4}, which correlates
with the granular structure revealed by the film electron microscopy,
Fig.\ref{fig1}.

Finally, the thermo-magnetic instabilities found in our study indicate an impeded heat flow and removal in these films. While QIS applications of thin films do not use an external magnetic
field, the transmons are driven at high frequencies and show elevated dissipation due to morphological inhomogeneities. In particular, the granular structure may contribute to their additional distortion of the current flow, and extra resistance \cite{Ramiere2022,Mizuno1984}. Another possible source is Kapitza resistance due to phonon flow mismatch at the boundaries and the interfaces \cite{Mizuno1984}. Finally, occasional vortices may enter the structure due to extreme demagnetization and sharp edges. With or without vortices, the heat dissipation and cooling may become a serious issue if the device is located in a high vacuum and can only
sink heat through a substrate.

On the bright side, the films show clean exponential attenuation of the weak magnetic field, signaling a fairly clean superconducting gap. Therefore, they still have the potential for improvement, and here we suggest some strategies.

\subsection{Suggested mitigation strategies}

A few suggestions can be made regarding applying such films for superconducting transmons. (1) To investigate the impeded heat transfer and dissipation, one may try to measure the coherence time
by comparing the same transmon in vacuum and immersed in liquid helium directly inside a mixing chamber of a dilution refrigerator. (2) Reducing the substrate thickness may help to improve
heat sinking when transmons are in a vacuum. (3) The most problematic
aspect of these films is that their granular structure may be fixed by switching
to a different deposition method. The alternative methods producing
better Nb films have been known since the 1980s \cite{Oya1986,Wagner1998}.
For example, molecular beam epitaxy (MBE) was shown to produce Nb
films of excellent crystallinity with $RRR\approx200$ and, remarkably,
$T_{c}=9.45$ K \cite{Oya1986}. In this case, different substrates,
better matching Nb thermal contraction and expansion, such as $\alpha-$Al$_{2}$O$_{3}$
and MgO were used. (4) One may also try to perform post-manufacturing
treatment of the prepared films, for example, using recently introduced
electro-annealing that was used for the optimization of quantum interference
devices \cite{Collienne2021}. (5) Finally, we recently
showed \cite{p-irr-bridge} that the introduction of true point-like
defects using proton irradiation expectedly reduces the $T_{c}$ somewhat,
to around 9.16 K, indicating that it affects the grains' interior.
However, it seems to also inhibit thermo-magnetic instabilities, perhaps
because these defects increase critical current density, thus reducing
vortex motion. Uniformly distributed,
dilute point-like defects may be beneficial for the transmons for other reasons. For
example, they may reduce (pin or alter otherwise) the activity of two-level systems (TLS)
and also prevent phase slips in the Josephson network of grain boundaries.

\section{Conclusions}

In summary, a comprehensive characterization of 160 nm thick Nb films
on {[}001{]} Si substrate used in superconducting transmons is presented.
Electron microscopy, transport, magnetization, quasiparticle spectroscopy, and real-space magneto-optical images all show that while these films have outstanding superconducting transition temperature of $T_{c}=9.35$ K and clean superconducting gap,
their behavior in the magnetic field is complicated exhibiting significantly
irreversible behavior, and insufficient heat conductance leading to
thermo-magnetic instabilities. These may present an issue for further
improvement of transmon quantum coherence. Possible mitigation strategies
are suggested, and the techniques presented here provide guidelines for comprehensive characterization that may be used to improve materials for superconducting quantum computing.

\begin{acknowledgments}
We thank V. Kogan, M. Zarea, A. Romanenko, and A. Grasselino for
discussions and attention. We thank the Rigetti fabrication team for
developing and manufacturing the studied films. This work
was supported by the U.S. Department of Energy, Office of Science,
National Quantum Information Science Research Centers, Superconducting
Quantum Materials and Systems Center (SQMS) under contract number
DE-AC02-07CH11359. The research was performed at the Ames National Laboratory,
operated for the U.S. DOE by Iowa State University under
contract \# DE-AC02-07CH11358.
\end{acknowledgments}




%

\end{document}